\documentclass[reprint,NumberedRefs]{JASA}
\usepackage{multirow}
\newcolumntype{P}[1]{>{\centering\arraybackslash}p{#1}}
\usepackage{ltablex}
\usepackage{graphics}
\usepackage{adjustbox}
\usepackage{footnotehyper}
\usepackage{tipa}
\usepackage{tikz,siunitx}
\usetikzlibrary{matrix,shadings,calc}
\usetikzlibrary{patterns}
\tikzset{ 
table/.style={
  nodes={rectangle,align=center},
  nodes in empty cells
}
}

\begin{document}

\title[JASA]{Automated Sex Classification of Children’s Voices and Changes in Differentiating Factors with Age}
\author{Fuling Chen}
\email{21552223@student.uwa.edu.au}
\author{Roberto Togneri}
\affiliation{Dept. of Electrical, Electronic and Computer Engineering, University of Western Australia, Perth, Western Australia, 6155, Australia}

\author{Murray Maybery}
\author{Diana Weiting Tan}
\altaffiliation{Also at: Macquarie School of Education, Macquarie University, Sydney, New South Wales, 2109, Australia.}
\affiliation{School of Psychological Science, University of Western Australia, Perth, Western Australia, 6155, Australia}


\date{\today} 

\begin{abstract}
Sex classification of children’s voices allows for an investigation of the development of secondary sex characteristics which has been a key interest in the field of speech analysis. This research investigated a broad range of acoustic features from scripted and spontaneous speech and applied a hierarchical clustering-based machine learning model to distinguish the sex of children aged between 5 and 15 years. We proposed an optimal feature set and our modelling achieved an average F1 score (the harmonic mean of the precision and recall) of 0.84 across all ages. Our results suggest that the sex classification is generally more accurate when a model is developed for each year group rather than for children in 4-year age bands, with classification accuracy being better for older age groups. We found that spontaneous speech could provide more helpful cues in sex classification than scripted speech, especially for children younger than 7 years. For younger age groups, a broad range of acoustic factors contributed evenly to sex classification, while for older age groups, F0-related acoustic factors were found to be the most critical predictors generally. Other important acoustic factors for older age groups include vocal tract length estimators, spectral flux, loudness and unvoiced features.
\end{abstract}


\maketitle

\section{\label{sec:1} Introduction}
The past two  decades  have witnessed great success in automated sex classification \footnote{The term “sex classification” refers to the binary classification of males and females. We acknowledge while there are different sex identities, including males, females, and other common identities such as LGBTIQA+ (lesbian, gay, bisexual, transgender, intersex, queer/questioning, asexual and other identities), previous works have focused on differences between males and females based on their self-identification with reference to sex as assigned at birth. We adopt this approach given the nature of the available substantial dataset, and also because sex as assigned at birth is a dominant factor in the development of the vocal system.} using speech which aims to identify the binary sex (male or female) of speakers by using computational modelling \citep{archana2015gender,childers1988automatic,ramdinmawii2016gender,zeng2006robust}. Automated sex classification assists in the study of the development of secondary sex characteristics which may be associated with health status \citep{daniel1980secondary,harlan1979secondary} and the presence of certain developmental disorders in children\citep{tordjman1997androgenic,baron2020foetal,worley2002secondary}. Sex classification has been found to be highly accurate for adult speakers. By contrast, the sex classification of young prepubescent child speakers is much less accurate \citep{shue2008role,chen2010using}. Automated sex classification from speech signals for children remains a challenge because most acoustic features are not highly distinguishable  between boys and girls. Additionally, another impediment to research in this space is that collecting children’s voices is time consuming and resource intensive. In the present study, we aim to develop a machine learning model for sex classification of children's voices and characterise the differentiating factors with age.

Numerous studies mainly used statistical analysis to investigate sex differences for individual acoustic features, and reported several features to be statistically significantly different between groups of boys and girls. The most typically studied acoustic features are fundamental frequency (F0) \citep{glaze1988acoustic,sorenson1989fundamental,sussman1994articulatory,whiteside2000some}, the first three formant frequencies (F1, F2 and F3, which will be referred to as Fn from here on) \citep{perry2001acoustic,vorperian2007vowel,whiteside2000some}, acoustic perturbation features, including jitter, shimmer and Harmonic-to-Noise Ratio (HNR) \citep{ferrand2000harmonics,guzman2014acoustic,sussman1994articulatory}, and the vocal tract length (VTL) estimator \citep{lehman2016estimation}. However, we have limited knowledge of whether these acoustic features are sufficient for sex classification for children, and whether the utility of the different acoustic features in classification varies with age.

Some studies have developed machine learning models to address the sex classification problem for children and achieved promising results. Such studies deployed several low-level descriptors, such as the commonly used 57 mel-frequency cepstral coefficients (19 static MFCCs, 19 $\Delta$MFCCs and 19 $\Delta^2$MFCCs) \citep{safavi2018automatic,sarma2020multi}, the Computational Paralinguistics Challenge baseline set ComParE (6373 features) \citep{kaya2017emotion}, the extended Geneva Minimalistic Acoustic Parameter Set (eGeMAPS, with 88 features) \citep{cummins2017enhancing,shaqra2019recognizing} and the linear predictive cepstral coefficients (LPCCs, with 13 features) \citep{ramteke2018gender}. However, such large numbers of low-level descriptors with severe multicollinearity and undefined physical meanings are hard to interpret, especially the cepstral-based features. Moreover, it is not clear which descriptors are discriminant in sex classification. The reason could be that the models were designed for sex classification only, but not considering the characterisation of the contributing factors. Additionally, these feature sets do not cover all acoustic features that could discriminate boys and girls as mentioned earlier, such as the VTL estimators.

Furthermore, it is known that children’s voice attributes are age dependent \citep{tavares2010normative}. Therefore, any successful models of sex classification in children are likely to be sensitive to the speaker’s age. What is of specific interest is the age at which a specific voice attribute shows a sex difference. The most important factor that influences children’s voice attributes is puberty \citep{abitbol1999sex,harries1997changes}, which usually starts between ages 8 and 13 years in girls and between ages 9 and 15 years in boys \citep{marshall1970variations,wheeler1991physical}. Despite such wide age ranges for puberty onset, each child experiences his/her puberty in a short period of time. Children tend to present with different voice quality before, during and after their puberty. Some studies investigated the variation of acoustic features with age in children of individual years \citep{vorperian2007vowel}; while other studies grouped children by their age, but with different aggregations (e.g., 5-8 years, 9-12 years and 13-15 years in \citep{safavi2018automatic}; 4-7 years, 8-11 years, 12-16 years in \citep{perry2001acoustic}; 8-9 years, 10-11 years, 12-13 years, 14-15 years in \citep{shue2008role}). It is not clear whether sex classification in children is any better when based on aggregation of year groups rather than based simply on each age year. 

There has been limited investigation of sex classification for children that has compared their scripted speech and spontaneous speech. Most studies of sex differences have used scripted productions \citep{ferrand2000harmonics,glaze1988acoustic,guzman2014acoustic,lehman2016estimation,perry2001acoustic,safavi2018automatic,sorenson1989fundamental,sussman1994articulatory,vorperian2007vowel,whiteside2000some}, spontaneous speech \citep{kaya2017emotion,safavi2018automatic,sorenson1989fundamental}, or both scripted and spontaneous speech combined into one cohort \citep{safavi2018automatic,sorenson1989fundamental}.  However, it was shown in several studies that significant differences existed between scripted and spontaneous speech in terms of acoustic and linguistic parameters \citep{farantouri2008linguistic,howell1991comparison,nakamura2008differences}. Accordingly, it is of interest to know more about the utility of scripted and spontaneous speech in sex classification of children and in mapping variation in distinguishing features across age. 

Additionally, as mentioned above, broad sets of features were largely popularly used in sex classification by means of machine learning models \citep{kaya2017emotion,ramteke2018gender,safavi2018automatic,zourmand2013gender}. However, severe multicollinearity could occur in these large feature sets. For example the mean value and the percentile values of the same acoustic source are likely to be strongly correlated. Few studies have taken multicollinearity into account and by not doing so limit the capability to properly characterise the important contributing factors in sex classification. 

Lastly, several studies have shown that some acoustic features could be used in differentiating the two sexes in children, but the particular features identified are not consistent across studies. For example, some studies reported girls had significantly higher F0 than boys \citep{glaze1988acoustic,whiteside2000some}, but other studies reported no difference between boys and girls in F0 \citep{sorenson1989fundamental,sussman1994articulatory}. Glaze et al. \citep{glaze1988acoustic} investigated F0 on the sustained neutral /a/ vowel; Whiteside and Hodgeson \citep{whiteside2000some} analysed F0 for the entire duration of the phrase-final vowel /a:/ following /b/, /k/ and /d$\mathfrak{z}$/; and Sussman and Sapienza \citep{sussman1994articulatory} analysed F0 on the sustained vowels /a/ /i/ and /u/. Whiteside and Hodgeson \citep{whiteside2000some} also reported that girls generally had higher Fn than boys across all ages. More specifically, Perry et al. \citep{perry2001acoustic} established that Fn differentiated sexes for children aged 4 years and above, while F0 differentiated sexes after 12 years of age. A systematic review \citep{vorperian2007vowel} showed age- and sex- related changes in F0 and Fn. Vorperian and Kent found a gradual reduction in Fn and the standard deviation of Fn (Fn std) with age, the emergence of male-female differences in Fn by age 4 years with more obvious differences by 8 years, and jumps in Fn at ages corresponding to growth spurts of the VTL. The acoustic perturbation features, including jitter, shimmer and HNR, have also been investigated for children in previous studies. For instance, it was found that significant sex differences in jitter were evident for adults, but  not for children \citep{sussman1994articulatory}. Ferrand \citep{ferrand2000harmonics} reported that girls presented significantly higher HNRs than boys. But another study \citep{guzman2014acoustic} investigating F0, Fn and perturbation features reported no significant sex differences for most of the acoustic features, except for shimmer and F3 for specific vowels. Thus, as alluded to above, although previous research has identified some acoustic cues that may discriminate boys’ and girls’ voices, outcomes are not consistent. 

The contributions of the present study are three-fold. Firstly, we identify the optimal feature set for voice sex classification in children, by comparing classification performance using different feature sets. These features include the acoustic features (abbreviated as AF from here on, including F0, Fn, VTL estimators and voice perturbation features), the extended Geneva Minimalistic Acoustic Parameter Set (eGeMAPS) \citep{eyben2015geneva} which covers most features in AF except for VTL estimators, and the combined feature set of eGeMAPS and VTL. Secondly, we establish the optimal age grouping strategy for sex classification by building models for children of each year and children of larger age bands (5-8 years, 9-12 years, and 13-15 years) and comparing the models’ performances. Thirdly, we use a correlation-matrix-based hierarchical clustering method applied on an optimal set of features to address multicollinearity while retaining the critical information. Lastly, we identify the critical factors in sex classification as they change across the childhood age range. 

\section{\label{sec:2} Dataset}
The OGI Kids’ Speech Corpus \citep{shobaki2000ogi}  is a collection of spontaneous and scripted speech recorded in the Northwest Regional School District near Portland, Oregon. A sex-balanced group of approximately 100 children per grade from Kindergarten (5 years) through to grade 10 (15 years), approximately 1100 children in total, participated in the collection. The protocol consisted of scripted speech and spontaneous speech. The scripted speech is composed of three types of stimulus; 205 isolated words, 100 prompted sentences, and 10 numeric strings. In the present study, each stimulus was regarded as one sample of scripted speech. All scripted speech utterances were rated on a three-point scale consisting of “good” utterances (the word is clearly intelligible with no significant background noise or extraneous speech), “questionable” (intelligible but accompanied by other sounds) or “bad” utterances (unintelligible or wrong word spoken). We only included good utterances in the present study. After the collection of scripted speech was completed, the experimenter asked the participants a series of questions intended to elicit spontaneous speech (e.g., “Tell me about your favourite movie”). In the present study, each recording of the spontaneous speech was segmented into 5 second timeframe samples. The total amount of speech recorded per subject was approximately 8-10 minutes. A summary of the number of participants and number of samples of scripted and spontaneous speech is shown, per year, in TABLE \ref{tab:dataset}.

\begin{table}[b]
\caption{Number of subjects and samples per year.}

\label{tab:dataset}
\begin{tabular}{P{0.08\linewidth}|P{0.12\linewidth}P{0.12\linewidth}|P{0.12\linewidth}P{0.12\linewidth}|P{0.12\linewidth}P{0.12\linewidth}}
\hline\hline
\multirow{2}{*}{age} & \multicolumn{2}{c|}{subjects} & \multicolumn{2}{c|}{scripted samples} & \multicolumn{2}{c}{spontaneous samples} \\ \cline{2-7} 
        & girls              & boys              & girls                 & boys                 & girls                  & boys                  \\ \hline
5       & 49                 & 39                & 1393                  & 1092                 & 1133                   & 925                   \\
6       & 31                 & 58                & 1585                  & 3131                 & 668 & 1190                  \\
7       & 61                 & 53                & 3237                  & 2685                 & 1367                   & 1184                  \\
8       & 52                 & 63                & 3262                  & 3726                 & 1223                   & 1410                  \\
9       & 44                 & 47                & 2508                  & 2642                 & 844 & 877                   \\
10      & 50                 & 49                & 2972                  & 2902                 & 957 & 952                   \\
11      & 55                 & 57                & 2001                  & 1992                 & 1004                   & 1097                  \\
12      & 51                 & 46                & 1872                  & 1754                 & 958 & 917                   \\
13      & 50                 & 49                & 1923                  & 1808                 & 930 & 910                   \\
14      & 39                 & 69                & 1586                  & 2773                 & 596 & 1171                  \\
15      & 29                 & 75                & 1132                  & 2898                 & 486 & 1388 \\                
\hline\hline
\end{tabular}
\end{table}

\section{\label{sec:3} Methodology}
\subsection{\label{sec:3:1} Feature sets}
\subsubsection{\label{sec:3:1:1} Acoustic features}
A set of 23 widely known AF were used to describe the vocal characteristics of each sample. Among these features, mean value of F0 (F0), standard deviation of F0 (F0 std), HNR, all jitter features (local Jitter, local absolute Jitter, rap Jitter, ppq5 Jitter and ddp Jitter) and all shimmer features (local Shimmer, apq3/5/11 Shimmer and dda Shimmer) were obtained from Parselmouth 0.3.3 which is a Python package for the Praat Software \citep{jadoul2018introducing}. The mean values of F1, F2, F3 and F4 measure the corresponding formants at each glottal pulse using the formant position formula \citep{puts2012masculine}. The apparent VTL was estimated using six features based on F1, F2, F3 and F4, namely formant dispersion (fdisp) \citep{fitch1997vocal}, average formant frequency (avgF) \citep{pisanski2011prioritization}, geometric mean formant frequency (mff) \citep{smith2005interaction}, Fitch formant estimate (fitch\_vtl) \citep{fitch1997vocal}, formant spacing ($\Delta$f) \citep{reby2003anatomical} and formant position (pF) \citep{puts2012masculine}.

\subsubsection{\label{sec:3:1:2} Extended Geneva Minimalistic Acoustic Parameter Set}
The extended Geneva Minimalistic Acoustic Parameter Set (eGeMAPS) \citep{eyben2015geneva} is composed of 88 low-level descriptors   which cover frequency related parameters (F0, Jitter, Fn frequencies and bandwidths), energy and amplitude related parameters (shimmer, loudness and HNR), spectral parameters (Alpha Ratio, Hammarberg Index, Spectral Slope, Fn energy, Harmonic difference H1-H2, H1-A3) and temporal features (rate of loudness peaks, voiced and unvoiced regions, and the number of continuous voiced regions per second). The parameters were derived using the default configuration in the openSMILE Python package \citep{eyben2010opensmile}.

\subsubsection{\label{sec:3:1:3} eGeMAPS+VTL}
Considering that eGeMAPS covers most features in the AF feature set except for the estimates of vocal tract length which could be important features in discriminating the two sexes, we proposed an extended feature set called eGeMAPS+VTL. eGeMAPS+VTL comprises 94 features that include the 88 low-level descriptors in eGeMAPS and the 6 estimators of VTL from the AF feature set, namely pF, fdisp, avgF, mff, fitch\_vtl and $\Delta$f.

\subsection{\label{sec:3:2} Acoustic characterisation (correlation-matrix-based hierarchical clustering)}
The correlation matrices of the feature sets for the children aged 5 to 15 years showed that some features were highly and consistently correlated throughout all ages (e.g., all jitter features in AF, all shimmer features in AF, and Fn bandwidths in eGeMAPS), and some features were highly correlated for younger children but were independent for older children (e.g., F0 related features were strongly correlated with Fn bandwidths in children aged 5 to 7 years, but became independent for children aged above 8 years). Such complexity of inter-correlations and severe multicollinearity introduces challenges in interpreting the contributing factors in sex classification.

Feature selection and variable clustering are the most popular methods for tackling the multicollinearity problem. Feature selection aims to select the most related feature among a group of ``similar” features. It is suitable for features that are known to be replaceable by other ``similar” features in the same group. In the present work, due to the complex inter-relationships among multiple features and the uniqueness of each feature, feature selection is not suited. Therefore, we used a correlation-matrix-based hierarchical clustering method to group the strongly correlated features and represent them with newly generated variables.

The method is one of the agglomerative clustering methods, which are popular in multiple fields \citep{gu2010study,liu2012correlation}. The idea of such methods is to group variables according to their ``similarity”. In the present work, we defined ``similarity” as the absolute value of the Pearson’s correlation coefficient of the two objects, which was also used for the multicollinearity diagnostic. The most similar clusters are merged at each step to produce larger clusters. The outcome of this clustering method is the distribution of the factors and their inclusive features. It should be noticed that some factors could be composed of multiple features, and some factors are just the features themselves, depending on the independency of the features. In the following sections, we will use the term ``factor” to refer to the independent cluster generated by the clustering algorithm, and the term ``feature” to refer to the original feature that was extracted as described in the Section \ref{sec:3:1}. The workflow of the clustering method and the details of the algorithm we implemented can be found in the Appendix.

\subsection{\label{sec:3:3} Extreme Random Forest Model}
The Extreme Random Forest (ERF) model is one of the most popular machine learning algorithms used for classification. It is an extension of the Random Forest (RF) model, and is an ensemble machine learning algorithm \citep{geurts2006extremely}. The ERF algorithm works by creating a large number of unpruned decision trees from the training dataset. In the case of classification, predictions are made by using majority voting. The ERF provides good predictive performance and low over-fitting by selecting a split point at random and fitting each decision tree on the whole training set.

Furthermore, the ERF provides factor  importance by computing the impurity of each node. The more a factor decreases the impurity, the more important the factor is. The impurity decreases provided by factors  can be averaged across trees to determine the factor importance. Factors that are selected at the top of trees are in general more important than factors that are selected at the end nodes of trees, as top splits lead to bigger information gains.

\subsection{\label{sec:3:4} Experiments and evaluation}
We conducted two steps of experimentation. In the first step, three sets of ERF models were trained by using the three sets of features (AF, eGeMAPS and eGeMAPS+VTL) with no application of the clustering method. This modelling was initially on all speech samples including scripted speech and spontaneous speech. In each set of models, we built 11 models by using samples of children per year, and 3 models by using samples of children per age band (5-8 years, 9-12 years, and 13-15 years), giving a total of 14 models per feature set. Depending on the performance of these three sets of models, two decisions were made to proceed to the next step: determining the optimal feature set, and whether one model per year or one model per age band is more appropriate for sex classification in children. In the second step, two sets of ERF models were trained by using the optimal feature set and the optimal age division with the application of the clustering method on scripted speech and spontaneous speech. Based on the outcomes of the second step, we then obtained the weights of the factors in distinguishing the male and female voices in each age division.

Because the samples of the two classes (boys and girls) are not strictly balanced for size, we applied Borderline-SMOTE \citep{han2005borderline} from the Python imblearn package to oversample the minor class to balance with the major class. 

The most suitable hyper-parameters of each ERF model were obtained by exhaustive search over specified parameter values and using a cross-validation splitting strategy of five folds with five repetitions, evaluated by the weighted F1 score. The F1 score is a measure of a test’s accuracy in statistical analysis of binary classification, which is calculated as in Eqs (\ref{eq:1}), (\ref{eq:2}), and (\ref{eq:3}). The final sex prediction of each subject was decided by majority voting of the predictions across all boy/girl samples from the test sets. 
\begin{footnotesize}
\begin{align}
 F1 &= 2\ddagger\frac{precision \ddagger recall}{precision+recall} \label{eq:1}\\ 
 precision &= \frac{true positive}{true positive+false positive} \label{eq:2}\\
 recall &= \frac{true positive}{true positive+false negative}
 \label{eq:3}
\end{align}
\end{footnotesize}

To evaluate the results of sex classification performance, we evaluated the F1 scores for both classes and the mean of the F1 scores of the two classes for the overall performance of each model. We also compared our results with the previous studies of sex classification using the same cohort \citep{safavi2018automatic,sarma2020multi}. 

\section{\label{sec:4} Results and discussion}
\subsection{\label{sec:4:1} Sex classification by using AF, eGeMAPS and eGeMAPS+VTL}

TABLE \ref{tab:result4.1} presents the F1 scores for girls and boys and the mean F1  scores for both classes  per age year and per age band, when using each of the three feature sets described in Section \ref{sec:3:1} (AF, eGeMAPS, and eGeMAPS+VTL). 

\begin{table}[b]
\caption{F1 scores for girls (G) and boys (B) and the mean F1 scores for both classes (Mean). The feature set sizes are presented in parentheses.}

\label{tab:result4.1}
\begin{adjustbox}{width=\linewidth}
\begin{tabular}{P{0.09\linewidth}|P{0.08\linewidth}P{0.08\linewidth}P{0.11\linewidth}|P{0.08\linewidth}P{0.08\linewidth}P{0.11\linewidth}|P{0.08\linewidth}P{0.08\linewidth}P{0.11\linewidth}}

\hline\hline
\multirow{2}{*}{age} & \multicolumn{3}{c|}{AF(23)} & \multicolumn{3}{c|}{eGeMAPS(88)} & \multicolumn{3}{c}{eGeMAPS +VTL(94)} \\ \cline{2-10} 
                     & G        & B       & Mean    & G      & B      & Mean            & G       & B       & Mean             \\ \hline
5                    & 0.68     & 0.63    & 0.66    & 0.87   & 0.82   & \textbf{0.85}   & 0.86    & 0.82    & 0.84             \\
6                    & 0.57     & 0.84    & 0.7     & 0.72   & 0.89   & \textbf{0.81}   & 0.69    & 0.88    & 0.79             \\
7                    & 0.72     & 0.57    & 0.64    & 0.73   & 0.69   & 0.71            & 0.75    & 0.7     & \textbf{0.73}    \\
8                    & 0.58     & 0.62    & 0.6     & 0.71   & 0.77   & 0.74            & 0.75    & 0.8     & \textbf{0.77}    \\
9                    & 0.79     & 0.81    & 0.8     & 0.79   & 0.81   & \textbf{0.8}    & 0.8     & 0.8     & \textbf{0.8}     \\
10                   & 0.67     & 0.67    & 0.67    & 0.78   & 0.77   & 0.78            & 0.8     & 0.78    & \textbf{0.79}    \\
11                   & 0.78     & 0.75    & 0.77    & 0.82   & 0.82   & 0.82            & 0.83    & 0.83    & \textbf{0.83}    \\
12                   & 0.81     & 0.77    & 0.79    & 0.85   & 0.82   & 0.83            & 0.89    & 0.86    & \textbf{0.87}    \\
13                   & 0.91     & 0.9     & 0.91    & 0.94   & 0.93   & \textbf{0.94}   & 0.94    & 0.93    & \textbf{0.94}    \\
14                   & 0.82     & 0.9     & 0.86    & 0.83   & 0.91   & 0.87            & 0.85    & 0.92    & \textbf{0.89}    \\
15                   & 0.74     & 0.88    & 0.81    & 0.95   & 0.98   & \textbf{0.96}   & 0.95    & 0.98    & \textbf{0.96}    \\\hline
5-8                  & 0.61     & 0.66    & 0.64    & 0.67   & 0.73   & 0.7             & 0.68    & 0.73    & 0.71             \\
9-12                 & 0.71     & 0.7     & 0.7     & 0.77   & 0.76   & 0.77            & 0.77    & 0.75    & 0.76             \\
13-15                & 0.76     & 0.81    & 0.79    & 0.81   & 0.85   & 0.83            & 0.82    & 0.86    & 0.84 \\                
\hline\hline
\end{tabular}
\end{adjustbox}
\end{table}

In comparing the performance of using AF, eGeMAPS, and eGeMAPS+VTL, eGeMAPS+VTL is generally more accurate in differentiating boys from girls than the other two feature sets. This result indicates the importance of the VTL features in sex classification for children, which is consistent with the findings in \citep{lehman2016estimation}. Our results also show that the eGeMAPS feature set provided more helpful information in sex classification than the AF set across all ages.

As expected, sex classification is more successful in older children than in younger children. Interestingly, we found that the F1 scores were higher (i.e., sex classification was more successful) for the analyses conducted for each year of age rather than for the broader age bands. The likely reason is that some acoustic factors vary with children’s growth, but at different rates for boys and girls. So the two sexes could experience similar changes in voice, but over different age ranges. For instance, Vorperian and Kent \citep{vorperian2007vowel} showed that the average F1-F2 acoustic space for boys at 4-6 years old largely overlapped with the acoustic space for girls at 5-8 years, with a similar pattern also observed for boys at 7-11 years and girls at 11-16 years. By mixing children of different ages, younger boys or older girls could be misclassified because of their similar acoustic characteristics. Therefore, we strongly suggest, given plenty of data across the age range, that children’s voice sex classification should be conducted based on each year, instead of broader age bands.

Sex classification has been investigated in the recent studies \citep{safavi2018automatic,sarma2020multi}, which used the same stimulus cohort but different features and models. In \citep{safavi2018automatic}, the researchers used cepstral features (57 MFCCs), and fed them into a Gaussian Mixture Model with a Universal Background Model (GMM-UBM) which achieved an accuracy of 0.77 for 5-8 year olds, 0.84 for 9-12 year olds, and 0.76 for 13-15 year olds. sex. And in \citep{sarma2020multi}, a Deep Neural Network (DNN) model  was applied on 60 MFCCs to the whole samples of children which resulted in an overall accuracy  of 0.82 . It is worth pointing out two limitations of the studies \citep{safavi2018automatic,sarma2020multi}. Firstly, the MFCC features are difficult to interpret in terms of human voice quality , therefore there is no evident acoustic analysis of how MFCC features could be different in sex for children. In our present work, we used the more explainable feature set eGeMAPS which could be used to describe voice quality in multiple ways. Secondly, the GMM-UBM model and the DNN model showed great capability in classification tasks, but these models  do not provide evidence on how they   make the classification decision with regard to the input data (e.g., factor importance).  While the models in these two studies provide rates of classification accuracy comparable to the rates reported for the model used in the present study, output from the models in \citep{safavi2018automatic,sarma2020multi} is not as amenable to interpretation as the output from the present model.

As discussed above, due to the best performance being given by the combination feature set of eGeMAPS+VTL based on each year, in the following sections, we will only consider using eGeMAPS+VTL features for children of each year.

\subsection{\label{sec:4:2} Sex classification of different types of speech}

TABLE \ref{tab:result4.2} presents the F1 scores for girls and boys and the average F1 scores across the two classes, when conducting analyses separately for scripted and spontaneous speech.

\begin{table}[b]
\caption{F1 scores for girls (G) and boys (B) and the macro average F1 scores for both classes (Mean), using scripted and spontaneous speech.}

\label{tab:result4.2}
\begin{tabular}{P{0.1\linewidth}|P{0.1\linewidth}P{0.1\linewidth}P{0.2\linewidth}|P{0.1\linewidth}P{0.1\linewidth}P{0.2\linewidth}}

\hline\hline
\multirow{2}{*}{age} & \multicolumn{3}{c|}{scripted} & \multicolumn{3}{c}{spontaneous} \\ \cline{2-7} 
 & G     & B     & Mean          & G      & B      & Mean           \\ \hline
5                    & 0.82  & 0.76  & 0.79          & 0.86   & 0.81   & \textbf{0.84}  \\
6                    & 0.69  & 0.88  & 0.79          & 0.74   & 0.86   & \textbf{0.8}   \\
7                    & 0.73  & 0.69  & 0.71          & 0.82   & 0.8    & \textbf{0.81}  \\
8                    & 0.73  & 0.79  & \textbf{0.76} & 0.73   & 0.79   & \textbf{0.76}  \\
9                    & 0.83  & 0.84  & \textbf{0.84} & 0.83   & 0.84   & 0.83           \\
10                   & 0.8   & 0.79  & \textbf{0.8}  & 0.79   & 0.8    & \textbf{0.8}   \\
11                   & 0.74  & 0.73  & 0.73          & 0.87   & 0.88   & \textbf{0.87}  \\
12                   & 0.87  & 0.83  & \textbf{0.85} & 0.87   & 0.84   & \textbf{0.85}  \\
13                   & 0.94  & 0.93  & \textbf{0.94} & 0.91   & 0.89   & 0.9            \\
14                   & 0.85  & 0.92  & 0.89          & 0.9    & 0.94   & \textbf{0.92}  \\
15                   & 0.95  & 0.98  & 0.96          & 0.96   & 0.99   & \textbf{0.98}  \\                
\hline\hline
\end{tabular}
\end{table}

Generally, the model  using spontaneous speech generated better performances than the models using the scripted speech. Few  studies have investigated the acoustic differences between scripted and spontaneous speech, which means there is a lack of knowledge about the salient acoustic factors that could be used to classify children by sex when using spontaneous speech rather than scripted speech. This could be due to several reasons, such as small samples, inconsistent data size between scripted speech and spontaneous speech, and weak methods in analysing the discriminant acoustic factors.

\subsection{\label{sec:4:3} Sex classification using individual features and clustered features}

TABLE \ref{tab:result4.3} presents the results of F1 scores for classification before clustering (BC-Mean) and after clustering (AC-Mean), by using scripted and spontaneous speech samples. The results in the “BC-Mean” columns in TABLE \ref{tab:result4.3} are the same as the results in the “Mean” columns in TABLE \ref{tab:result4.2}, which were obtained by using the individual 94 eGeMAPS+VTL features, while the results in the “AC-Mean” columns in TABLE \ref{tab:result4.3} were obtained by applying the clustering algorithm. Generally, after clustering, the F1 scores were slightly lower than the F1 scores before clustering, which is within expectation. This is because, after clustering, data dimensionality was reduced considerably from the original 94 dimensions to between 41 and 63 dimensions (shown as column “n” in TABLE \ref{tab:result4.3}), with n varying with the age of the children and the type of speech. After clustering, the redundant information carried by highly correlated features was eliminated, and the essential information was condensed within each independent cluster. 

\begin{table}[b]
\caption{The F1 scores before clustering (BC-Mean) and after clustering (AC-Mean), by using scripted and spontaneous speech samples.}

\label{tab:result4.3}
\begin{tabular}{P{0.09\linewidth}|P{0.15\linewidth}P{0.15\linewidth}P{0.08\linewidth}|P{0.15\linewidth}P{0.15\linewidth}P{0.08\linewidth}}

\hline\hline
\multirow{2}{*}{age} & \multicolumn{3}{c|}{scripted} & \multicolumn{3}{c}{spontaneous} \\ \cline{2-7} 
 & BC-Mean     & AC-Mean     & n\footnotemark[1]          & BC-Mean      & AC-Mean     & n           \\ \hline
5                    & 0.79  & 0.73           & 54   & 0.84   & \textbf{0.74}   & 46    \\
6                    & 0.79  & 0.77           & 56   & 0.8    & \textbf{0.78}   & 50    \\
7                    & 0.71  & 0.66           & 55   & 0.81   & \textbf{0.76}   & 44    \\
8                    & 0.76  & \textbf{0.71}  & 60   & 0.76   & 0.68            & 47    \\
9                    & 0.84  & 0.75           & 59   & 0.83   & \textbf{0.8}    & 41    \\
10                   & 0.8   & \textbf{0.79}  & 58   & 0.8    & 0.73            & 48    \\
11                   & 0.73  & 0.77           & 59   & 0.87   & \textbf{0.86}   & 50    \\
12                   & 0.85  & 0.8            & 59   & 0.85   & \textbf{0.85}   & 45    \\
13                   & 0.94  & \textbf{0.94}  & 58   & 0.9    & 0.92            & 49    \\
14                   & 0.89  & 0.86           & 63   & 0.92   & \textbf{0.88}   & 51    \\
15                   & 0.96  & 0.94           & 59   & 0.98   & \textbf{0.96}   & 49    \\                
\hline\hline
\end{tabular}
\footnotetext[1]{n is the number factors generated after clustering.}
\end{table}

As the F1 scores after clustering are not significantly less than the F1 scores before clustering, we regarded these independent clusters suitable for further investigation of the salient acoustic factors in sex classification. 

\subsection{\label{sec:4:4} Critical factors}
In summary of Sections \ref{sec:4:1}, \ref{sec:4:2} and \ref{sec:4:3}, the sex classification model benefits more from using spontaneous speech for children of each individual year, and the sex classification performance is acceptable after clustering. In this section, we present and discuss the important factors derived from the modelling outlined in Section \ref{sec:4:3} for spontaneous speech. 

We took the clusters generated for the results in the column ``AC-Mean” in TABLE \ref{tab:result4.3}, and referring to the clusters as acoustic factors, then extracted the factor importance of these clusters from the ERF models. The factor importance was presented in the form of the weight which was the percentage of the contribution of the factor.     

TABLE \ref{tab:clustertop10} presents the clustering pattern and the weights of the top 10 important factors across all ages. The features column lists the eGeMAPS+VTL features which constitute the clustering factors. The remaining columns provide the weights for each year across the listed features. For each age year, the 11 weights shown (the top 10 shown in percentage and the last row) sum up to 100\%. The features which are grouped in the same cluster are marked with the same symbol and their cells merged where possible (e.g., the cluster marked with ``$\ddagger$" includes all features from F0 mean to UVL for most age years, with the exception of ages 13 and 14 where only the first 5 features are included). The weight associated with the cluster is provided in the merged cells. For completeness, the acoustic factors not in the top 10 for some age years are also included but without their weights (e.g., the cluster marked with ``$\ddagger$" is not in the top 10 for ages 8 and 10). We excluded features that were not among the top factors of any age year and included them in the last row labeled "not in top 10" with their aggregated weight. More detailed information about TABLE \ref{tab:clustertop10} can be found in the supplementary material\footnote{See supplementary material at https://github.com/FulingChen/Automated-Sex-Classification-of-Children-Voices-and-Changes-in-Differentiating-Factors-with-Age/blob/main/SuppPub1.xlsx for the clustering pattern and the weights of the factors across all ages}, which includes all the 94 original features of eGeMAPS+VTL.

\begin{tiny}
\begin{center}
\tabcolsep=0pt
\renewcommand{\arraystretch}{1}
\begin{longtable}{|l|l|l|l|l|l|l|l|l|l|l|l|}

\caption{The clustering pattern and the weights of the top 10 important factors across all ages. } \label{tab:clustertop10} \\
\hline\hline  & \multicolumn{11}{c|}{Weights of each year}\\ \cline{2-12} 
\multirow{-2}{*}{features} & \multicolumn{1}{c|}{5}     & \multicolumn{1}{c|}{6}     & \multicolumn{1}{c|}{7}     & \multicolumn{1}{c|}{8}     & \multicolumn{1}{c|}{9}     & \multicolumn{1}{c|}{10}    & \multicolumn{1}{c|}{11}    & \multicolumn{1}{c|}{12}     & \multicolumn{1}{c|}{13}     & \multicolumn{1}{c|}{14}     & 15    \\ \hline

\endfirsthead
    \multicolumn{11}{c}
    {\tablename\ \thetable\ -- \textit{Continued}} \\
\hline \multicolumn{12}{|c|}{Weights of each year}\\ \cline{2-12} 
\multirow{-2}{*}{features} & \multicolumn{1}{c|}{5}     & \multicolumn{1}{c|}{6}     & \multicolumn{1}{c|}{7}     & \multicolumn{1}{c|}{8}     & \multicolumn{1}{c|}{9}     & \multicolumn{1}{c|}{10}    & \multicolumn{1}{c|}{11}    & \multicolumn{1}{c|}{12}     & \multicolumn{1}{c|}{13}     & \multicolumn{1}{c|}{14}     & 15    \\ \hline
 
\endhead
\hline
\multicolumn{12}{l}{\it(Continued)}
\endfoot
\hline \hline
\endlastfoot
F0 mean & \multicolumn{1}{c|}{$\ddagger$} & \multicolumn{1}{c|}{$\ddagger$} & \multicolumn{1}{c|}{$\ddagger$} & \multicolumn{1}{c|}{$\ddagger$} & \multicolumn{1}{c|}{$\ddagger$} & \multicolumn{1}{c|}{$\ddagger$} & \multicolumn{1}{c|}{$\ddagger$} & \multicolumn{1}{c|}{$\ddagger$} & \multicolumn{1}{c|}{$\ddagger$} & \multicolumn{1}{c|}{$\ddagger$} & $\ddagger$ \\ \cline{1-1}
F0 p20 & \multicolumn{1}{c|}{$\ddagger$} & \multicolumn{1}{c|}{$\ddagger$} & \multicolumn{1}{c|}{$\ddagger$} & \multicolumn{1}{c|}{$\ddagger$} & \multicolumn{1}{c|}{$\ddagger$} & \multicolumn{1}{c|}{$\ddagger$} & \multicolumn{1}{c|}{$\ddagger$} & \multicolumn{1}{c|}{$\ddagger$} & \multicolumn{1}{c|}{$\ddagger$} & \multicolumn{1}{c|}{$\ddagger$} & $\ddagger$ \\ \cline{1-1}
F0 p50 & \multicolumn{1}{c|}{$\ddagger$} & \multicolumn{1}{c|}{$\ddagger$} & \multicolumn{1}{c|}{$\ddagger$} & \multicolumn{1}{c|}{$\ddagger$} & \multicolumn{1}{c|}{$\ddagger$} & \multicolumn{1}{c|}{$\ddagger$} & \multicolumn{1}{c|}{4\%} & \multicolumn{1}{c|}{10\%} & \multicolumn{1}{c|}{31\%} & \multicolumn{1}{c|}{14\%} & 6\%\\ \cline{1-1}
F0 p80 & \multicolumn{1}{c|}{$\ddagger$} & \multicolumn{1}{c|}{$\ddagger$} & \multicolumn{1}{c|}{$\ddagger$} & \multicolumn{1}{c|}{$\ddagger$} & \multicolumn{1}{c|}{$\ddagger$} & \multicolumn{1}{c|}{$\ddagger$} & \multicolumn{1}{c|}{$\ddagger$} & \multicolumn{1}{c|}{$\ddagger$} & \multicolumn{1}{c|}{$\ddagger$} & \multicolumn{1}{c|}{$\ddagger$} & $\ddagger$ \\ \cline{1-1}
HNR & \multicolumn{1}{c|}{$\ddagger$} & \multicolumn{1}{c|}{$\ddagger$} & \multicolumn{1}{c|}{$\ddagger$} & \multicolumn{1}{c|}{$\ddagger$} & \multicolumn{1}{c|}{$\ddagger$} & \multicolumn{1}{c|}{$\ddagger$} & \multicolumn{1}{c|}{$\ddagger$} & \multicolumn{1}{c|}{$\ddagger$} & \multicolumn{1}{c|}{$\ddagger$} & \multicolumn{1}{c|}{$\ddagger$} & $\ddagger$ \\ \cline{1-1} \cline{7-12} 
H1-A3 & \multicolumn{1}{c|}{$\ddagger$} & \multicolumn{1}{c|}{$\ddagger$} & \multicolumn{1}{c|}{$\ddagger$} & \multicolumn{1}{c|}{$\ddagger$} & \multicolumn{1}{c|}{$\ddagger$} & \multicolumn{1}{c|}{$||$} & \multicolumn{1}{c|}{$||$} & \multicolumn{1}{c|}{$||$} & \multicolumn{1}{c|}{$||$} & \multicolumn{1}{c|}{$||$} & $||$ \\ \cline{1-1}
mfcc1V & \multicolumn{1}{c|}{$\ddagger$} & \multicolumn{1}{c|}{$\ddagger$} & \multicolumn{1}{c|}{$\ddagger$} & \multicolumn{1}{c|}{$\ddagger$} & \multicolumn{1}{c|}{$\ddagger$} & \multicolumn{1}{c|}{7\%} & \multicolumn{1}{c|}{5\%} & \multicolumn{1}{c|}{5\%} & \multicolumn{1}{c|}{6\%} & \multicolumn{1}{c|}{6\%} & 3\% \\ \cline{1-1} \cline{7-12} 
F1 & \multicolumn{1}{c|}{$\ddagger$} & \multicolumn{1}{c|}{$\ddagger$} & \multicolumn{1}{c|}{$\ddagger$} & \multicolumn{1}{c|}{$\ddagger$} & \multicolumn{1}{c|}{$\ddagger$} & \multicolumn{1}{c|}{$\ddagger$} & \multicolumn{1}{c|}{$\ddagger$} & \multicolumn{1}{c|}{$\ddagger$} & \multicolumn{1}{c|}{$\bullet$} & \multicolumn{1}{c|}{$\bullet$} & $\ddagger$ \\ \cline{1-1}
F2 & \multicolumn{1}{c|}{6\%} & \multicolumn{1}{c|}{4\%} & \multicolumn{1}{c|}{3\%} & \multicolumn{1}{c|}{$\ddagger$} & \multicolumn{1}{c|}{4\%} & \multicolumn{1}{c|}{$\ddagger$} & \multicolumn{1}{c|}{$\ddagger$} & \multicolumn{1}{c|}{$\ddagger$} & \multicolumn{1}{c|}{$\bullet$} & \multicolumn{1}{c|}{$\bullet$} & $\ddagger$ \\ \cline{1-1}
F3 & \multicolumn{1}{c|}{$\ddagger$} & \multicolumn{1}{c|}{$\ddagger$} & \multicolumn{1}{c|}{$\ddagger$} & \multicolumn{1}{c|}{$\ddagger$} & \multicolumn{1}{c|}{$\ddagger$} & \multicolumn{1}{c|}{$\ddagger$} & \multicolumn{1}{c|}{$\ddagger$} & \multicolumn{1}{c|}{$\ddagger$} & \multicolumn{1}{c|}{$\bullet$} & \multicolumn{1}{c|}{$\bullet$} & $\ddagger$ \\ \cline{1-1}
F1 BD & \multicolumn{1}{c|}{$\ddagger$} & \multicolumn{1}{c|}{$\ddagger$} & \multicolumn{1}{c|}{$\ddagger$} & \multicolumn{1}{c|}{$\ddagger$} & \multicolumn{1}{c|}{$\ddagger$} & \multicolumn{1}{c|}{$\ddagger$} & \multicolumn{1}{c|}{$\ddagger$} & \multicolumn{1}{c|}{$\ddagger$} & \multicolumn{1}{c|}{$\bullet$} & \multicolumn{1}{c|}{$\bullet$} & $\ddagger$ \\ \cline{1-1}
F2 BD & \multicolumn{1}{c|}{$\ddagger$} & \multicolumn{1}{c|}{$\ddagger$} & \multicolumn{1}{c|}{$\ddagger$} & \multicolumn{1}{c|}{$\ddagger$} & \multicolumn{1}{c|}{$\ddagger$} & \multicolumn{1}{c|}{$\ddagger$} & \multicolumn{1}{c|}{$\ddagger$} & \multicolumn{1}{c|}{$\ddagger$} & \multicolumn{1}{c|}{$\bullet$} & \multicolumn{1}{c|}{3\%} & $\ddagger$ \\ \cline{1-1}
F3 BD & \multicolumn{1}{c|}{$\ddagger$} & \multicolumn{1}{c|}{$\ddagger$} & \multicolumn{1}{c|}{$\ddagger$} & \multicolumn{1}{c|}{$\ddagger$} & \multicolumn{1}{c|}{$\ddagger$} & \multicolumn{1}{c|}{$\ddagger$} & \multicolumn{1}{c|}{$\ddagger$} & \multicolumn{1}{c|}{$\ddagger$} & \multicolumn{1}{c|}{$\bullet$} & \multicolumn{1}{c|}{$\bullet$} & $\ddagger$ \\ \cline{1-1}
aR V & \multicolumn{1}{c|}{$\ddagger$} & \multicolumn{1}{c|}{$\ddagger$} & \multicolumn{1}{c|}{$\ddagger$} & \multicolumn{1}{c|}{$\ddagger$} & \multicolumn{1}{c|}{$\ddagger$} & \multicolumn{1}{c|}{$\ddagger$} & \multicolumn{1}{c|}{$\ddagger$} & \multicolumn{1}{c|}{$\ddagger$} & \multicolumn{1}{c|}{$\bullet$} & \multicolumn{1}{c|}{$\bullet$} & $\ddagger$ \\ \cline{1-1}
hI V & \multicolumn{1}{c|}{$\ddagger$} & \multicolumn{1}{c|}{$\ddagger$} & \multicolumn{1}{c|}{$\ddagger$} & \multicolumn{1}{c|}{$\ddagger$} & \multicolumn{1}{c|}{$\ddagger$} & \multicolumn{1}{c|}{$\ddagger$} & \multicolumn{1}{c|}{$\ddagger$} & \multicolumn{1}{c|}{$\ddagger$} & \multicolumn{1}{c|}{$\bullet$} & \multicolumn{1}{c|}{$\bullet$} & $\ddagger$ \\ \cline{1-1} \cline{3-3} \cline{8-11}
F1 std & \multicolumn{1}{c|}{$\ddagger$} & \multicolumn{1}{c|}{$\diamond$} & \multicolumn{1}{c|}{$\ddagger$} & \multicolumn{1}{c|}{$\ddagger$} & \multicolumn{1}{c|}{$\ddagger$} & \multicolumn{1}{c|}{$\ddagger$} & \multicolumn{1}{c|}{$\diamond$} & \multicolumn{1}{c|}{$\diamond$} & \multicolumn{1}{c|}{$\diamond$} & \multicolumn{1}{c|}{$\diamond$} & $\ddagger$ \\ \cline{1-1}
F2 std & \multicolumn{1}{c|}{$\ddagger$} & \multicolumn{1}{c|}{3\%} & \multicolumn{1}{c|}{$\ddagger$} & \multicolumn{1}{c|}{$\ddagger$} & \multicolumn{1}{c|}{$\ddagger$} & \multicolumn{1}{c|}{$\ddagger$} & \multicolumn{1}{c|}{$\diamond$} & \multicolumn{1}{c|}{3\%} & \multicolumn{1}{c|}{$\diamond$} & \multicolumn{1}{c|}{$\diamond$} & $\ddagger$ \\ \cline{1-1}
F3 std & \multicolumn{1}{c|}{$\ddagger$} & \multicolumn{1}{c|}{$\diamond$} & \multicolumn{1}{c|}{$\ddagger$} & \multicolumn{1}{c|}{$\ddagger$} & \multicolumn{1}{c|}{$\ddagger$} & \multicolumn{1}{c|}{$\ddagger$} & \multicolumn{1}{c|}{$\diamond$} & \multicolumn{1}{c|}{$\diamond$} & \multicolumn{1}{c|}{$\diamond$} & \multicolumn{1}{c|}{$\diamond$} & $\ddagger$ \\ \cline{1-1} \cline{3-3} \cline{7-11}
UVL & \multicolumn{1}{c|}{$\ddagger$} & \multicolumn{1}{c|}{$\ddagger$} & \multicolumn{1}{c|}{$\ddagger$} & \multicolumn{1}{c|}{$\ddagger$} & \multicolumn{1}{c|}{$\ddagger$} & \multicolumn{1}{c|}{} & \multicolumn{1}{c|}{} & \multicolumn{1}{c|}{} & \multicolumn{1}{c|}{} & \multicolumn{1}{c|}{} & $\ddagger$ \\ \hline
slpV0-500 & \multicolumn{1}{c|}{} & \multicolumn{1}{c|}{3\%} & \multicolumn{1}{c|}{3\%} & \multicolumn{1}{c|}{3\%} & \multicolumn{1}{c|}{} & \multicolumn{1}{c|}{} & \multicolumn{1}{c|}{} & \multicolumn{1}{c|}{$\ddagger$} & \multicolumn{1}{c|}{} & \multicolumn{1}{c|}{4\%} &  \\ \hline
F1 BD std & \multicolumn{1}{c|}{} & \multicolumn{1}{c|}{} & \multicolumn{1}{c|}{$\ddagger$} & \multicolumn{1}{c|}{} & \multicolumn{1}{c|}{} & \multicolumn{1}{c|}{} & \multicolumn{1}{c|}{} & \multicolumn{1}{c|}{} & \multicolumn{1}{c|}{} & \multicolumn{1}{c|}{} & \\
\hline
avgF & \multicolumn{1}{c|}{\dotfill} & \multicolumn{1}{c|}{\dotfill} & \multicolumn{1}{c|}{\dotfill} & \multicolumn{1}{c|}{\dotfill} & \multicolumn{1}{c|}{\dotfill} & \multicolumn{1}{c|}{\dotfill} & \multicolumn{1}{c|}{\dotfill} & \multicolumn{1}{c|}{\dotfill} & \multicolumn{1}{c|}{\dotfill} & \multicolumn{1}{c|}{\dotfill} & \dotfill \\ \cline{1-1}
$\Delta$f & \multicolumn{1}{c|}{\dotfill} & \multicolumn{1}{c|}{\dotfill} & \multicolumn{1}{c|}{\dotfill} & \multicolumn{1}{c|}{\dotfill} & \multicolumn{1}{c|}{\dotfill} & \multicolumn{1}{c|}{\dotfill} & \multicolumn{1}{c|}{\dotfill} & \multicolumn{1}{c|}{\dotfill} & \multicolumn{1}{c|}{\dotfill} & \multicolumn{1}{c|}{\dotfill} & \dotfill \\ \cline{1-1}
pF & \multicolumn{1}{c|}{5\%} & \multicolumn{1}{c|}{4\%} & \multicolumn{1}{c|}{3\%} & \multicolumn{1}{c|}{\dotfill} & \multicolumn{1}{c|}{\dotfill} & \multicolumn{1}{c|}{7\%} & \multicolumn{1}{c|}{\dotfill} & \multicolumn{1}{c|}{4\%} & \multicolumn{1}{c|}{7\%} & \multicolumn{1}{c|}{4\%} & 7\% \\ \cline{1-1}
mff & \multicolumn{1}{c|}{\dotfill} & \multicolumn{1}{c|}{\dotfill} & \multicolumn{1}{c|}{\dotfill} & \multicolumn{1}{c|}{\dotfill} & \multicolumn{1}{c|}{\dotfill} & \multicolumn{1}{c|}{\dotfill} & \multicolumn{1}{c|}{\dotfill} & \multicolumn{1}{c|}{\dotfill} & \multicolumn{1}{c|}{\dotfill} & \multicolumn{1}{c|}{\dotfill} & \dotfill \\ \cline{1-1}
fitch\_vtl & \multicolumn{1}{c|}{\dotfill} & \multicolumn{1}{c|}{\dotfill} & \multicolumn{1}{c|}{\dotfill} & \multicolumn{1}{c|}{\dotfill} & \multicolumn{1}{c|}{\dotfill} & \multicolumn{1}{c|}{\dotfill} & \multicolumn{1}{c|}{\dotfill} & \multicolumn{1}{c|}{\dotfill} & \multicolumn{1}{c|}{\dotfill} & \multicolumn{1}{c|}{\dotfill} & \dotfill \\ \hline
aR UV & \multicolumn{1}{c|}{\hrulefill} & \multicolumn{1}{c|}{\hrulefill} & \multicolumn{1}{c|}{\hrulefill} & \multicolumn{1}{c|}{\hrulefill} & \multicolumn{1}{c|}{\hrulefill} & \multicolumn{1}{c|}{\hrulefill} & \multicolumn{1}{c|}{\hrulefill} & \multicolumn{1}{c|}{\hrulefill} & \multicolumn{1}{c|}{\hrulefill} & \multicolumn{1}{c|}{\hrulefill} & \hrulefill \\ \cline{1-1}
hI UV & \multicolumn{1}{c|}{3\%} & \multicolumn{1}{c|}{\hrulefill} & \multicolumn{1}{c|}{\hrulefill} & \multicolumn{1}{c|}{3\%} & \multicolumn{1}{c|}{4\%} & \multicolumn{1}{c|}{\hrulefill} & \multicolumn{1}{c|}{3\%} & \multicolumn{1}{c|}{3\%} & \multicolumn{1}{c|}{\hrulefill} & \multicolumn{1}{c|}{5\%} & 5\% \\ \cline{1-1}
slpUV0-500 & \multicolumn{1}{c|}{\hrulefill} & \multicolumn{1}{c|}{\hrulefill} & \multicolumn{1}{c|}{\hrulefill} & \multicolumn{1}{c|}{\hrulefill} & \multicolumn{1}{c|}{\hrulefill} & \multicolumn{1}{c|}{\hrulefill} & \multicolumn{1}{c|}{\hrulefill} & \multicolumn{1}{c|}{\hrulefill} & \multicolumn{1}{c|}{\hrulefill} & \multicolumn{1}{c|}{\hrulefill} & \hrulefill \\ \hline
SF mean & \multicolumn{1}{c|}{//} & \multicolumn{1}{c|}{//} & \multicolumn{1}{c|}{//} & \multicolumn{1}{c|}{//} & \multicolumn{1}{c|}{//} & \multicolumn{1}{c|}{//} & \multicolumn{1}{c|}{//} & \multicolumn{1}{c|}{//} & \multicolumn{1}{c|}{//} & \multicolumn{1}{c|}{//} & // \\ \cline{1-1}
SF UV & \multicolumn{1}{c|}{//} & \multicolumn{1}{c|}{//} & \multicolumn{1}{c|}{//} & \multicolumn{1}{c|}{3\%} & \multicolumn{1}{c|}{4\%} & \multicolumn{1}{c|}{//} & \multicolumn{1}{c|}{//} & \multicolumn{1}{c|}{//} & \multicolumn{1}{c|}{//} & \multicolumn{1}{c|}{//} & // \\ \cline{1-1}
SF V & \multicolumn{1}{c|}{//} & \multicolumn{1}{c|}{//} & \multicolumn{1}{c|}{//} & \multicolumn{1}{c|}{//} & \multicolumn{1}{c|}{//} & \multicolumn{1}{c|}{//} & \multicolumn{1}{c|}{//} & \multicolumn{1}{c|}{//} & \multicolumn{1}{c|}{//} & \multicolumn{1}{c|}{//} & // \\ \cline{1-6} \cline{8-8} \cline{11-11}
LD mean & \multicolumn{1}{c|}{$\star$} & \multicolumn{1}{c|}{$\star$} & \multicolumn{1}{c|}{$\star$} & \multicolumn{1}{c|}{$\star$} & \multicolumn{1}{c|}{$\star$} & \multicolumn{1}{c|}{//} & \multicolumn{1}{c|}{$\star$} & \multicolumn{1}{c|}{//} & \multicolumn{1}{c|}{//} & \multicolumn{1}{c|}{$\star$} & // \\ \cline{1-1}
LD p20 & \multicolumn{1}{c|}{$\star$} & \multicolumn{1}{c|}{$\star$} & \multicolumn{1}{c|}{$\star$} & \multicolumn{1}{c|}{$\star$} & \multicolumn{1}{c|}{$\star$} & \multicolumn{1}{c|}{//} & \multicolumn{1}{c|}{$\star$} & \multicolumn{1}{c|}{//} & \multicolumn{1}{c|}{//} & \multicolumn{1}{c|}{$\star$} & // \\ \cline{1-1}
LD p50 & \multicolumn{1}{c|}{$\star$} & \multicolumn{1}{c|}{$\star$} & \multicolumn{1}{c|}{$\star$} & \multicolumn{1}{c|}{$\star$} & \multicolumn{1}{c|}{$\star$} & \multicolumn{1}{c|}{//} & \multicolumn{1}{c|}{$\star$} & \multicolumn{1}{c|}{//} & \multicolumn{1}{c|}{//} & \multicolumn{1}{c|}{$\star$} & // \\ \cline{1-1}
LD p80 & \multicolumn{1}{c|}{3\%} & \multicolumn{1}{c|}{3\%} & \multicolumn{1}{c|}{4\%} & \multicolumn{1}{c|}{3\%} & \multicolumn{1}{c|}{6\%} & \multicolumn{1}{c|}{8\%} & \multicolumn{1}{c|}{$\star$} & \multicolumn{1}{c|}{4\%} & \multicolumn{1}{c|}{3\%} & \multicolumn{1}{c|}{$\star$} & 8\% \\ \cline{1-1}
SL & \multicolumn{1}{c|}{$\star$} & \multicolumn{1}{c|}{$\star$} & \multicolumn{1}{c|}{$\star$} & \multicolumn{1}{c|}{$\star$} & \multicolumn{1}{c|}{$\star$} & \multicolumn{1}{c|}{//} & \multicolumn{1}{c|}{$\star$} & \multicolumn{1}{c|}{//} & \multicolumn{1}{c|}{//} & \multicolumn{1}{c|}{$\star$} & // \\ \cline{1-1} \cline{3-4} \cline{8-8} \cline{11-12} 
slpUV500-1500 & \multicolumn{1}{c|}{$\star$} & \multicolumn{1}{c|}{} & \multicolumn{1}{c|}{} & \multicolumn{1}{c|}{$\star$} & \multicolumn{1}{c|}{$\star$} & \multicolumn{1}{c|}{//} & \multicolumn{1}{c|}{} & \multicolumn{1}{c|}{//} & \multicolumn{1}{c|}{//} & \multicolumn{1}{c|}{} &  \\ \cline{1-6} \cline{8-8} \cline{11-12} 
LD p02 & \multicolumn{1}{c|}{} & \multicolumn{1}{c|}{} & \multicolumn{1}{c|}{3\%} & \multicolumn{1}{c|}{//} & \multicolumn{1}{c|}{//} & \multicolumn{1}{c|}{//} & \multicolumn{1}{c|}{} & \multicolumn{1}{c|}{//} & \multicolumn{1}{c|}{//} & \multicolumn{1}{c|}{} & // \\ \cline{1-4} \cline{8-8} \cline{11-11}
LD rslp & \multicolumn{1}{c|}{//} & \multicolumn{1}{c|}{} & \multicolumn{1}{c|}{} & \multicolumn{1}{c|}{//} & \multicolumn{1}{c|}{//} & \multicolumn{1}{c|}{//} & \multicolumn{1}{c|}{} & \multicolumn{1}{c|}{//} & \multicolumn{1}{c|}{//} & \multicolumn{1}{c|}{} & // \\ \cline{1-1} \cline{3-4} \cline{8-8} \cline{11-11}
LD rslp std & \multicolumn{1}{c|}{//} & \multicolumn{1}{c|}{} & \multicolumn{1}{c|}{} & \multicolumn{1}{c|}{//} & \multicolumn{1}{c|}{//} & \multicolumn{1}{c|}{//} & \multicolumn{1}{c|}{} & \multicolumn{1}{c|}{//} & \multicolumn{1}{c|}{//} & \multicolumn{1}{c|}{} & // \\ \cline{1-4} \cline{8-8} \cline{11-11}
LD fslp & \multicolumn{1}{c|}{} & \multicolumn{1}{c|}{} & \multicolumn{1}{c|}{} & \multicolumn{1}{c|}{//} & \multicolumn{1}{c|}{//} & \multicolumn{1}{c|}{//} & \multicolumn{1}{c|}{} & \multicolumn{1}{c|}{//} & \multicolumn{1}{c|}{//} & \multicolumn{1}{c|}{} & // \\ \cline{1-4} \cline{8-8} \cline{11-11}
LD fslp std & \multicolumn{1}{c|}{} & \multicolumn{1}{c|}{} & \multicolumn{1}{c|}{} & \multicolumn{1}{c|}{//} & \multicolumn{1}{c|}{//} & \multicolumn{1}{c|}{//} & \multicolumn{1}{c|}{} & \multicolumn{1}{c|}{//} & \multicolumn{1}{c|}{//} & \multicolumn{1}{c|}{} & \multicolumn{1}{c|}{//} \\ \cline{1-6} \cline{8-8} \cline{10-12} 
mfcc1 & \multicolumn{1}{c|}{} & \multicolumn{1}{c|}{3\%} & \multicolumn{1}{c|}{} & \multicolumn{1}{c|}{} & \multicolumn{1}{c|}{4\%} & \multicolumn{1}{c|}{//} & \multicolumn{1}{c|}{} & \multicolumn{1}{c|}{//} & \multicolumn{1}{c|}{} & \multicolumn{1}{c|}{} & 5\% \\ \hline
F0 std & \multicolumn{1}{c|}{} & \multicolumn{1}{c|}{} & \multicolumn{1}{c|}{} & \multicolumn{1}{c|}{3\%} & \multicolumn{1}{c|}{} & \multicolumn{1}{c|}{} & \multicolumn{1}{c|}{3\%} & \multicolumn{1}{c|}{4\%} & \multicolumn{1}{c|}{} & \multicolumn{1}{c|}{} &  \\ \hline
LD std & \multicolumn{1}{c|}{+} & \multicolumn{1}{c|}{} & \multicolumn{1}{c|}{} & \multicolumn{1}{c|}{3\%} & \multicolumn{1}{c|}{+} & \multicolumn{1}{c|}{7\%} & \multicolumn{1}{c|}{} & \multicolumn{1}{c|}{} & \multicolumn{1}{c|}{} & \multicolumn{1}{c|}{} &  \\ \cline{1-5} \cline{7-12} 
mfcc1 std & \multicolumn{1}{c|}{\hrulefill} & \multicolumn{1}{c|}{} & \multicolumn{1}{c|}{} & \multicolumn{1}{c|}{} & \multicolumn{1}{c|}{+} & \multicolumn{1}{c|}{} & \multicolumn{1}{c|}{} & \multicolumn{1}{c|}{} & \multicolumn{1}{c|}{} & \multicolumn{1}{c|}{} &  \\ \cline{1-5} \cline{7-12} 
F2 BD std & \multicolumn{1}{c|}{+} & \multicolumn{1}{c|}{3\%} & \multicolumn{1}{c|}{} & \multicolumn{1}{c|}{} & \multicolumn{1}{c|}{+} & \multicolumn{1}{c|}{} & \multicolumn{1}{c|}{} & \multicolumn{1}{c|}{} & \multicolumn{1}{c|}{} & \multicolumn{1}{c|}{} &  \\ \cline{1-1} \cline{3-5} \cline{7-12} 
F3 BD std & \multicolumn{1}{c|}{4\%} & \multicolumn{1}{c|}{4\%} & \multicolumn{1}{c|}{} & \multicolumn{1}{c|}{} & \multicolumn{1}{c|}{4\%} & \multicolumn{1}{c|}{} & \multicolumn{1}{c|}{} & \multicolumn{1}{c|}{} & \multicolumn{1}{c|}{} & \multicolumn{1}{c|}{} &  \\ \cline{1-1} \cline{3-5} \cline{7-12} 
aR V std & \multicolumn{1}{c|}{+} & \multicolumn{1}{c|}{} & \multicolumn{1}{c|}{} & \multicolumn{1}{c|}{} & \multicolumn{1}{c|}{+} & \multicolumn{1}{c|}{5\%} & \multicolumn{1}{c|}{} & \multicolumn{1}{c|}{} & \multicolumn{1}{c|}{} & \multicolumn{1}{c|}{} &  \\ \cline{1-1} \cline{3-5} \cline{8-12} 
hI V std & \multicolumn{1}{c|}{+} & \multicolumn{1}{c|}{} & \multicolumn{1}{c|}{} & \multicolumn{1}{c|}{} & \multicolumn{1}{c|}{+} & \multicolumn{1}{c|}{+} & \multicolumn{1}{c|}{} & \multicolumn{1}{c|}{} & \multicolumn{1}{c|}{} & \multicolumn{1}{c|}{} &  \\ \cline{1-1} \cline{3-5} \cline{7-12} 
mfcc1V  std & \multicolumn{1}{c|}{+} & \multicolumn{1}{c|}{} & \multicolumn{1}{c|}{} & \multicolumn{1}{c|}{} & \multicolumn{1}{c|}{+} & \multicolumn{1}{c|}{} & \multicolumn{1}{c|}{} & \multicolumn{1}{c|}{} & \multicolumn{1}{c|}{} & \multicolumn{1}{c|}{} &  \\ \hline
F1amp & \multicolumn{1}{c|}{} & \multicolumn{1}{c|}{} & \multicolumn{1}{c|}{} & \multicolumn{1}{c|}{} & \multicolumn{1}{c|}{} & \multicolumn{1}{c|}{} & \multicolumn{1}{c|}{$\sim$} & \multicolumn{1}{c|}{} & \multicolumn{1}{c|}{} & \multicolumn{1}{c|}{} &  \\ \cline{1-7} \cline{9-12} 
F2amp & \multicolumn{1}{c|}{} & \multicolumn{1}{c|}{} & \multicolumn{1}{c|}{} & \multicolumn{1}{c|}{} & \multicolumn{1}{c|}{} & \multicolumn{1}{c|}{} & \multicolumn{1}{c|}{$\sim$} & \multicolumn{1}{c|}{} & \multicolumn{1}{c|}{} & \multicolumn{1}{c|}{} &  \\ \cline{1-7} \cline{9-12} 
F3amp & \multicolumn{1}{c|}{} & \multicolumn{1}{c|}{} & \multicolumn{1}{c|}{} & \multicolumn{1}{c|}{} & \multicolumn{1}{c|}{} & \multicolumn{1}{c|}{} & \multicolumn{1}{c|}{$\sim$} & \multicolumn{1}{c|}{} & \multicolumn{1}{c|}{} & \multicolumn{1}{c|}{} &  \\ \cline{1-7} \cline{9-12} 
F1amp std & \multicolumn{1}{c|}{} & \multicolumn{1}{c|}{} & \multicolumn{1}{c|}{} & \multicolumn{1}{c|}{} & \multicolumn{1}{c|}{} & \multicolumn{1}{c|}{} & \multicolumn{1}{c|}{3\%} & \multicolumn{1}{c|}{} & \multicolumn{1}{c|}{} & \multicolumn{1}{c|}{} &  \\ \cline{1-7} \cline{9-12} 
F2amp std & \multicolumn{1}{c|}{} & \multicolumn{1}{c|}{} & \multicolumn{1}{c|}{} & \multicolumn{1}{c|}{} & \multicolumn{1}{c|}{} & \multicolumn{1}{c|}{} & \multicolumn{1}{c|}{$\sim$} & \multicolumn{1}{c|}{} & \multicolumn{1}{c|}{} & \multicolumn{1}{c|}{} &  \\ \cline{1-7} \cline{9-12} 
F3amp std & \multicolumn{1}{c|}{} & \multicolumn{1}{c|}{} & \multicolumn{1}{c|}{} & \multicolumn{1}{c|}{} & \multicolumn{1}{c|}{} & \multicolumn{1}{c|}{} & \multicolumn{1}{c|}{$\sim$} & \multicolumn{1}{c|}{} & \multicolumn{1}{c|}{} & \multicolumn{1}{c|}{} &  \\ \cline{1-7} \cline{9-12} 
VPS & \multicolumn{1}{c|}{} & \multicolumn{1}{c|}{} & \multicolumn{1}{c|}{} & \multicolumn{1}{c|}{} & \multicolumn{1}{c|}{} & \multicolumn{1}{c|}{} & \multicolumn{1}{c|}{$\sim$} & \multicolumn{1}{c|}{} & \multicolumn{1}{c|}{} & \multicolumn{1}{c|}{} &  \\ \hline
mfcc2 & \multicolumn{1}{c|}{3\%} & \multicolumn{1}{c|}{} & \multicolumn{1}{c|}{} & \multicolumn{1}{c|}{} & \multicolumn{1}{c|}{4\%} & \multicolumn{1}{c|}{} & \multicolumn{1}{c|}{3\%} & \multicolumn{1}{c|}{} & \multicolumn{1}{c|}{} & \multicolumn{1}{c|}{3\%} &  \\ \cline{1-5} \cline{7-12} 
mfcc2V & \multicolumn{1}{c|}{3\%} & \multicolumn{1}{c|}{3\%} & \multicolumn{1}{c|}{5\%} & \multicolumn{1}{c|}{3\%} & \multicolumn{1}{c|}{$\triangleright$} & \multicolumn{1}{c|}{} & \multicolumn{1}{c|}{3\%} & \multicolumn{1}{c|}{} & \multicolumn{1}{c|}{} & \multicolumn{1}{c|}{3\%} &  \\ \hline
mfcc2 std & \multicolumn{1}{c|}{} & \multicolumn{1}{c|}{} & \multicolumn{1}{c|}{} & \multicolumn{1}{c|}{} & \multicolumn{1}{c|}{} & \multicolumn{1}{c|}{} & \multicolumn{1}{c|}{3\%} & \multicolumn{1}{c|}{} & \multicolumn{1}{c|}{} & \multicolumn{1}{c|}{} &  \\ \hline
mfcc3 & \multicolumn{1}{c|}{} & \multicolumn{1}{c|}{} & \multicolumn{1}{c|}{} & \multicolumn{1}{c|}{3\%} & \multicolumn{1}{c|}{3\%} & \multicolumn{1}{c|}{} & \multicolumn{1}{c|}{} & \multicolumn{1}{c|}{} & \multicolumn{1}{c|}{2\%} & \multicolumn{1}{c|}{} &  \\ \cline{1-5} \cline{7-9} \cline{11-12} 
mfcc3V & \multicolumn{1}{c|}{3\%} & \multicolumn{1}{c|}{} & \multicolumn{1}{c|}{3\%} & \multicolumn{1}{c|}{3\%} & \multicolumn{1}{c|}{$\triangleleft$} & \multicolumn{1}{c|}{} & \multicolumn{1}{c|}{} & \multicolumn{1}{c|}{} & \multicolumn{1}{c|}{$\triangleleft$} & \multicolumn{1}{c|}{} &  \\ \hline
mfcc3 std & \multicolumn{1}{c|}{} & \multicolumn{1}{c|}{} & \multicolumn{1}{c|}{} & \multicolumn{1}{c|}{} & \multicolumn{1}{c|}{} & \multicolumn{1}{c|}{5\%} & \multicolumn{1}{c|}{} & \multicolumn{1}{c|}{} & \multicolumn{1}{c|}{} & \multicolumn{1}{c|}{} &  \\ \hline
jitter std & \multicolumn{1}{c|}{} & \multicolumn{1}{c|}{} & \multicolumn{1}{c|}{} & \multicolumn{1}{c|}{} & \multicolumn{1}{c|}{} & \multicolumn{1}{c|}{} & \multicolumn{1}{c|}{} & \multicolumn{1}{c|}{} & \multicolumn{1}{c|}{6\%} & \multicolumn{1}{c|}{} &  \\ \hline
shimmer std & \multicolumn{1}{c|}{} & \multicolumn{1}{c|}{} & \multicolumn{1}{c|}{} & \multicolumn{1}{c|}{} & \multicolumn{1}{c|}{4\%} & \multicolumn{1}{c|}{} & \multicolumn{1}{c|}{} & \multicolumn{1}{c|}{} & \multicolumn{1}{c|}{} & \multicolumn{1}{c|}{} &  \\ \hline
HNR std & \multicolumn{1}{c|}{} & \multicolumn{1}{c|}{} & \multicolumn{1}{c|}{} & \multicolumn{1}{c|}{} & \multicolumn{1}{c|}{3\%} & \multicolumn{1}{c|}{4\%} & \multicolumn{1}{c|}{} & \multicolumn{1}{c|}{} & \multicolumn{1}{c|}{2\%} & \multicolumn{1}{c|}{} &  \\ \hline
H1-H2 & \multicolumn{1}{c|}{3\%} & \multicolumn{1}{c|}{} & \multicolumn{1}{c|}{3\%} & \multicolumn{1}{c|}{} & \multicolumn{1}{c|}{} & \multicolumn{1}{c|}{} & \multicolumn{1}{c|}{} & \multicolumn{1}{c|}{4\%} & \multicolumn{1}{c|}{2\%} & \multicolumn{1}{c|}{} & 3\% \\ \hline
H1-A3 std & \multicolumn{1}{c|}{} & \multicolumn{1}{c|}{} & \multicolumn{1}{c|}{} & \multicolumn{1}{c|}{} & \multicolumn{1}{c|}{} & \multicolumn{1}{c|}{} & \multicolumn{1}{c|}{4\%} & \multicolumn{1}{c|}{3\%} & \multicolumn{1}{c|}{5\%} & \multicolumn{1}{c|}{2\%} &  \\ \hline

slpV500-1500 & \multicolumn{1}{c|}{} & \multicolumn{1}{c|}{} & \multicolumn{1}{c|}{3\%} & \multicolumn{1}{c|}{} & \multicolumn{1}{c|}{} & \multicolumn{1}{c|}{7\%} & \multicolumn{1}{c|}{} & \multicolumn{1}{c|}{4\%} & \multicolumn{1}{c|}{3\%} & \multicolumn{1}{c|}{3\%} & 3\% \\ \hline
mfcc4V & \multicolumn{1}{c|}{} & \multicolumn{1}{c|}{2\%} & \multicolumn{1}{c|}{} & \multicolumn{1}{c|}{} & \multicolumn{1}{c|}{} & \multicolumn{1}{c|}{7\%} & \multicolumn{1}{c|}{} & \multicolumn{1}{c|}{} & \multicolumn{1}{c|}{} & \multicolumn{1}{c|}{} & 3\% \\ \hline
LDPPS & \multicolumn{1}{c|}{3\%} & \multicolumn{1}{c|}{} & \multicolumn{1}{c|}{} & \multicolumn{1}{c|}{} & \multicolumn{1}{c|}{} & \multicolumn{1}{c|}{4\%} & \multicolumn{1}{c|}{} & \multicolumn{1}{c|}{} & \multicolumn{1}{c|}{} & \multicolumn{1}{c|}{} & \multicolumn{1}{l|}{} \\ \hline
fdisp & \multicolumn{1}{c|}{} & \multicolumn{1}{c|}{} & \multicolumn{1}{c|}{3\%} & \multicolumn{1}{c|}{3\%} & \multicolumn{1}{c|}{} & \multicolumn{1}{c|}{} & \multicolumn{1}{c|}{3\%} & \multicolumn{1}{c|}{} & \multicolumn{1}{c|}{} & \multicolumn{1}{c|}{} & 5\% \\ \hline
not in top 10 & \multicolumn{1}{c|}{66\%} & \multicolumn{1}{c|}{67\%} & \multicolumn{1}{c|}{64\%} & \multicolumn{1}{c|}{70\%} & \multicolumn{1}{c|}{60\%} & \multicolumn{1}{c|}{39\%} & \multicolumn{1}{c|}{67\%} & \multicolumn{1}{c|}{56\%} & \multicolumn{1}{c|}{34\%} & \multicolumn{1}{c|}{54\%} & 51\% \\\hline
\end{longtable}
\end{center}
\end{tiny}
\normalsize
We found that, for children aged below 12 years old, there were no obvious acoustic factors that could significantly differentiate boys from girls, with all weights lower than 10\%. Each acoustic factor counts evenly in sex classification for young children, and the final judgements were made based on considering all the involved factors. As children’s age increased, some acoustic factors started to play important roles in sex classification.

The critical factors could vary with age. For instance, the 19 features (F0 mean, F0 p20, F0 p50, F0 p80, HNR, H1-A3, mfcc1V, F1, F2, F3, F1 BD, F2 BD, F3 BD, aR V, hI V, F1 std, F2 std, F3 std and UVL) formed a large cluster (marked with ``$\ddagger$") and became an important factor for young children aged between 6 to 9 years, as shown in TABLE \ref{tab:clustertop10}. However, with the increasing age, this big cluster started to break down into several small clusters, and some of these small clusters were identified to be important factors in sex classification. 

For those features that were consistently highly correlated and commonly appeared in the top 10 important factors across the age range, we identified them as critical acoustic factors and discussed them in the following sections. These important factors are 1) F0 and HNR (including F0 mean, F0 p20, F0 p50, F0 p80 and HNR), 2) Fn and Fn BD (including F1, F2, F3, F1 BD, F2 BD, F3 BD), 3) H1-A3 and mfcc1V, 4) VTL (including avgF, $\Delta$f, pF, mff and fitch\_vtl), 5) Spectral flux (including SF mean, SF UV and SF V), 6) unvoiced features (including aR UV, hI UV and slpUV0-500), and 7) loudness (including LD mean, LD p20, LD p50, LD p80 and SL). Considering the inclusive features are highly correlated (r$>$.75), in the following sections, we selected one or a few commonly discussed features to represent the cluster without loss of generality. We did not analyse the representatives generated by the clustering algorithm, because the application of PCA involved in the clustering algorithm on multiple features would affect the interpretation. We report the t-test statistic (t), p-value (p) and Cohen’s d (d) for each comparison. Due to the large data size, we relied on Cohen’s d (d) in drawing conclusions about the statistical analyses. Our interpretation is to refer to effect sizes as small ($\mid$d$\mid<$0.2), small to medium (0.2$\leq \mid$d$\mid<$0.5), medium to large (0.5$\leq \mid$d$\mid<$0.8, marked in bold), and large ($\mid$d$\mid \geq$0.8, marked in bold), based on benchmarks suggested by Cohen \citep{cohen2013statistical}.

\begin{table}[ht]
\caption{Summary based on t tests of differences on the features between boys and girls. (\#$>$0.05, †$<$0.05, ††$<$0.01, †††$<$0.001; negative values of d and t indicate that the mean value of the feature is larger in boys than in girls)}
\label{tab:stats}
\begin{scriptsize}
\begin{tabular}{cc|ccccccccccc}
\hline\hline
\multicolumn{2}{c|}{feature} & 5 & 6 & 7 & 8 & 9 & 10 & 11 & 12 & 13 & 14 & 15 \\ \hline
\multicolumn{1}{c|}{\multirow{3}{*}{\begin{tabular}[c]{@{}c@{}}F0\\ mean\end{tabular}}} & t & 6.3 & 1.2 & -3.0 & 1.2 & 7.8 & 6.6 & \textbf{11.9} & \textbf{19.9} & \textbf{45.9} & \textbf{23.7} & \textbf{22.1} \\
\multicolumn{1}{c|}{} & p & ††† & \# & †† & \# & ††† & ††† & ††† & ††† & ††† & ††† & ††† \\
\multicolumn{1}{c|}{} & d & 0.3 & 0.1 & -0.1 & 0.1 & 0.4 & 0.3 & \textbf{0.5} & \textbf{1.0} & \textbf{2.2} & \textbf{1.2} & \textbf{1.2} \\ \hline
\multicolumn{1}{c|}{\multirow{3}{*}{HNR}} & t & 5.4 & 0.3 & -2.1 & -0.2 & 6.6 & \textbf{12.1} & 7.9 & \textbf{19.6} & \textbf{42.4} & \textbf{27.6} & \textbf{24.8} \\
\multicolumn{1}{c|}{} & p & ††† & \# & † & \# & ††† & ††† & ††† & ††† & ††† & ††† & ††† \\
\multicolumn{1}{c|}{} & d & 0.3 & 0.0 & -0.1 & 0.0 & 0.3 & \textbf{0.6} & 0.4 & \textbf{0.9} & \textbf{2.0} & \textbf{1.4} & \textbf{1.3} \\ \hline
\multicolumn{1}{c|}{\multirow{3}{*}{F1}} & t & 4.8 & 2.4 & 2.3 & -1.5 & 3.7 & \textbf{11.6} & 8.9 & \textbf{14.0} & \textbf{15.4} & \textbf{14.4} & \textbf{13.9} \\
\multicolumn{1}{c|}{} & p & ††† & † & † & \# & ††† & ††† & ††† & ††† & ††† & ††† & ††† \\
\multicolumn{1}{c|}{} & d & 0.2 & 0.1 & 0.1 & -0.1 & 0.2 & \textbf{0.6} & 0.4 & \textbf{0.7} & \textbf{0.7} & \textbf{0.8} & \textbf{0.8} \\ \hline
\multicolumn{1}{c|}{\multirow{3}{*}{F2}} & t & 3.9 & 4.3 & 3.6 & 0.4 & 6.1 & \textbf{11.5} & 9.3 & \textbf{15.0} & \textbf{16.9} & \textbf{11.7} & \textbf{12.5} \\
\multicolumn{1}{c|}{} & p & ††† & ††† & ††† & \# & ††† & ††† & ††† & ††† & ††† & ††† & ††† \\
\multicolumn{1}{c|}{} & d & 0.2 & 0.2 & 0.2 & 0.0 & 0.3 & \textbf{0.5} & 0.4 & \textbf{0.7} & \textbf{0.8} & \textbf{0.6} & \textbf{0.7} \\ \hline
\multicolumn{1}{c|}{\multirow{3}{*}{F3}} & t & 3.9 & 4.0 & 1.4 & 0.7 & 5.2 & 9.4 & 9.6 & \textbf{14.3} & \textbf{15.8} & \textbf{10.9} & \textbf{13.1} \\
\multicolumn{1}{c|}{} & p & ††† & ††† & \# & \# & ††† & ††† & ††† & ††† & ††† & ††† & ††† \\
\multicolumn{1}{c|}{} & d & 0.2 & 0.2 & 0.1 & 0.0 & 0.3 & 0.4 & 0.4 & \textbf{0.7} & \textbf{0.8} & \textbf{0.6} & \textbf{0.7} \\ \hline
\multicolumn{1}{c|}{\multirow{3}{*}{\begin{tabular}[c]{@{}c@{}}H1-\\ A3\end{tabular}}} & t & 2.3 & 5.7 & 2.1 & 4.3 & -2.8 & -10.6 & \textbf{-11.8} & \textbf{-13.0} & \textbf{-25.7} & \textbf{-22.8} & -7.6 \\
\multicolumn{1}{c|}{} & p & † & ††† & † & ††† & †† & ††† & ††† & ††† & ††† & ††† & ††† \\
\multicolumn{1}{c|}{} & d & 0.1 & 0.3 & 0.1 & 0.2 & -0.1 & -0.5 & \textbf{-0.5} & \textbf{-0.6} & \textbf{-1.2} & \textbf{-1.2} & -0.4 \\ \hline
\multicolumn{1}{c|}{\multirow{3}{*}{\begin{tabular}[c]{@{}c@{}}fitch\\ \_vtl\end{tabular}}} & t & 0.5 & 7.0 & 2.2 & 5.3 & 0.0 & -13.4 & -2.6 & -8.5 & \textbf{-18.0} & \textbf{-15.0} & \textbf{-13.3} \\
\multicolumn{1}{c|}{} & p & \# & ††† & † & ††† & \# & ††† & †† & ††† & ††† & ††† & ††† \\
\multicolumn{1}{c|}{} & d & 0.0 & 0.3 & 0.1 & 0.2 & 0.0 & -0.6 & -0.1 & -0.4 & \textbf{-0.8} & \textbf{-0.8} & \textbf{-0.7} \\ \hline
\multicolumn{1}{c|}{\multirow{3}{*}{\begin{tabular}[c]{@{}c@{}}SF\\ mean\end{tabular}}} & t & 2.5 & 3.8 & -7.8 & -4.8 & -8.6 & \textbf{-13.1} & -1.4 & -8.6 & \textbf{-13.5} & -0.1 & 7.3 \\
\multicolumn{1}{c|}{} & p & † & ††† & ††† & ††† & ††† & ††† & \# & ††† & ††† & \# & ††† \\
\multicolumn{1}{c|}{} & d & 0.1 & 0.2 & -0.3 & -0.2 & -0.4 & \textbf{-0.6} & -0.1 & -0.4 & \textbf{-0.6} & 0.0 & 0.4 \\ \hline
\multicolumn{1}{c|}{\multirow{3}{*}{\begin{tabular}[c]{@{}c@{}}aR\_\\ UV\end{tabular}}} & t & -3.1 & 2.4 & 7.0 & 4.8 & -2.9 & 0.0 & 3.7 & -5.3 & -0.4 & \textbf{-15.5} & \textbf{-15.5} \\
\multicolumn{1}{c|}{} & p & †† & †† & ††† & ††† & †† & \# & ††† & ††† & \# & ††† & ††† \\
\multicolumn{1}{c|}{} & d & -0.1 & 0.1 & 0.3 & 0.2 & -0.1 & 0.0 & 0.2 & -0.2 & 0.0 & \textbf{-0.8} & \textbf{-0.8} \\ \hline
\multicolumn{1}{c|}{\multirow{3}{*}{\begin{tabular}[c]{@{}c@{}}LD\\ mean\end{tabular}}} & t & 2.0 & 0.9 & -9.4 & -8.1 & -7.5 & \textbf{-11.2} & 0.2 & -7.3 & -9.3 & 5.1 & \textbf{16.5} \\
\multicolumn{1}{c|}{} & p & † & \# & ††† & ††† & ††† & ††† & \# & ††† & ††† & ††† & ††† \\
\multicolumn{1}{c|}{} & d & 0.1 & 0.1 & -0.4 & -0.3 & -0.4 & \textbf{-0.5} & 0.0 & -0.3 & -0.4 & 0.3 & \textbf{0.9}
\\

\hline\hline
\end{tabular}

\end{scriptsize}


\end{table}

\begin{figure}[tp]
\figline{\hfill\fig{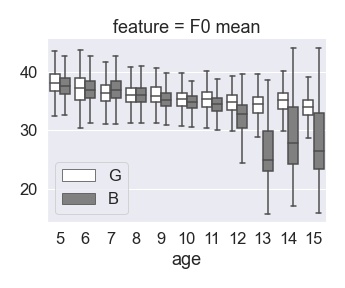}{.5\linewidth}{(a)}\label{fig:F0mean} 
\fig{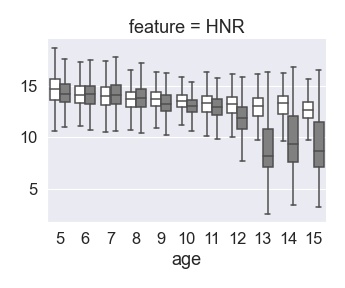}{.5\linewidth}{(b)}\label{fig:HNR}\hfill}

\figline{\hfill\fig{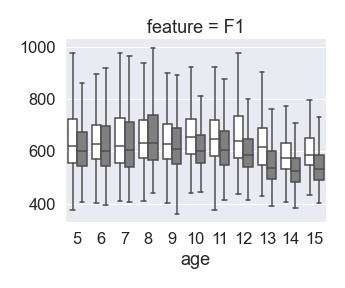}{.5\linewidth}{(c)}\label{fig:F1}
\fig{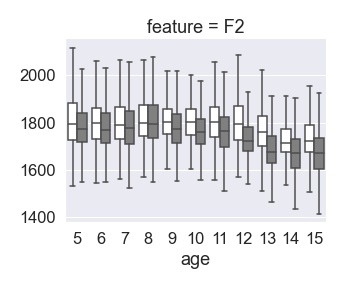}{.5\linewidth}{(d)}\label{fig:F2}\hfill}

\figline{\hfill\fig{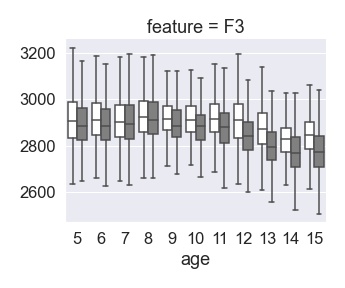}{.5\linewidth}{(e)}\label{fig:F3}
\fig{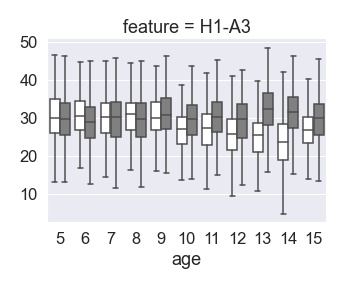}{.5\linewidth}{(f)}\label{fig:H1A3}\hfill}

\figline{\hfill\fig{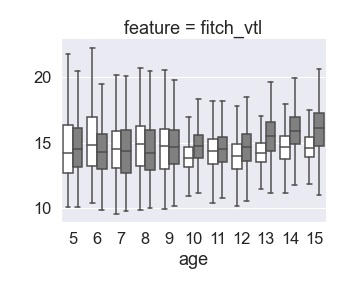}{.5\linewidth}{(g)}\label{fig:fitchvtl} \fig{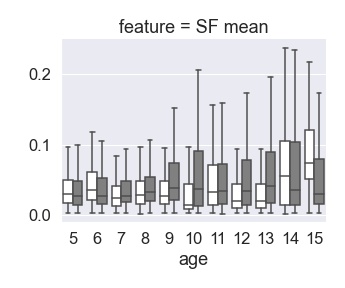}{.5\linewidth}{(h)}\label{fig:sfmean}\hfill}

\figline{\hfill\fig{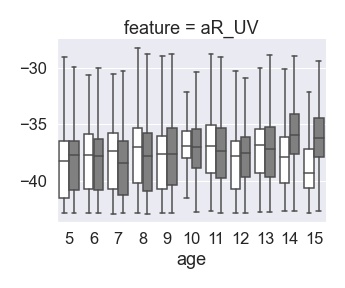}{.5\linewidth}{(i)}\label{fig:aruv}
\fig{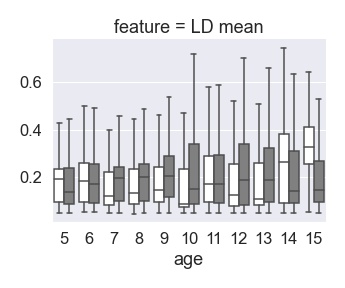}{.5\linewidth}{(j)}\label{fig:ldmean}\hfill}

\caption{Box plots of the features. (white - girls, grey - boys)}
\label{fig:boxplot}
\end{figure}

\subsubsection{\label{sec:4:4:1} F0 and HNR}
F0 mean, F0 p20, F0 p50 and F0 p80 (these 4 features are named F0 from here on) and HNR consistently appeared in the same factor across all ages. From TABLE \ref{tab:clustertop10}, it is observed that F0 and HNR did not contribute substantially to sex classification in younger children (weights less than 6\% for 5 to 11 year-olds) but started to be more discriminant for older children (weights 10\%-31\% in 12-, 13-, and 14-year-olds). As described earlier, for younger children, F0 and HNR, together with other acoustic features, formed a large cluster which consists of 19 features in total,  which gradually broke down into several small clusters and individual features with the increase in age. At ages 13 and 14, the cluster composed of F0 and HNR became independent of other features and was also the most important factor in sex classification.

Because the F0 and HNR related features are strongly correlated, we selected the F0 mean and HNR to represent the cluster and conducted the following analysis. FIG.\ref{fig:F0mean}, \ref{fig:HNR} and TABLE \ref{tab:stats} presents the F0 mean and HNR distributions and the statistical summary for both sexes across the age range from 5 to 15 years old. Specifically, the effect sizes for F0 and HNR are large in children older than 12, and are medium to large in 10- and 11-year-olds. F0 and HNR have small effects for children aged between 6 and 8 years. 

\subsubsection{\label{sec:4:4:2} Fn and Fn BD}
Fn and Fn BD consistently appeared in the same cluster throughout the age range as they were strongly correlated, and they were grouped with F0 and HNR for young children ($<$13 years old). As Fn and Fn BD were highly correlated, we selected Fn to represent their cluster. FIG.\ref{fig:F1}, \ref{fig:F2}, \ref{fig:F3} and TABLE \ref{tab:stats} show the Fn distributions and the statistical summary.

The sex differences in Fn and Fn BD and the variation with age have been rarely reported in developmental studies. In the present study, the results reveal that: (1) Fn reduces with advancing age in both sexes; (2) Fn is generally lower in boys than in girls across all ages, an outcome consistent with results reported by \citep{bennett1981vowel,kent2018static,vorperian2007vowel}, and (3) the differences in Fn between boys and girls become more consistent and more pronounced with increasing age, which is consistent with the findings in \citep{kent2018static,vorperian2007vowel}.  

It is known that Fn is related to the wave resonance of the pharynx \citep{fant1973speech}, which, together with changes associated with puberty \citep{fitch1999morphology}, explains why the sex differences in Fn increase with increasing age. Another possible reason for the sex differences in Fn are sex-specific articulatory behaviours. It was found that most vowels are articulated with considerable variation  in the size of the mouth opening, tongue dorsum position and jaw movements, and Fn is sensitive to such changes \citep{fant1970acoustic,ladefoged1979formant,stevens1955development}. For instance, males tend to lengthen their vocal tract by using a more pronounced degree of lip protrusion than females which could effectively decrease Fn \citep{stevens1955development}. It was also reported that the sex differences in F1 could be derived from the sex differences in the width and height of the lip opening and amount of jaw opening during the production of /æ/ \citep{lindblom1971acoustical,stevens1955development}. 

\subsubsection{\label{sec:4:4:3} H1-A3 and mfcc1V}
Because mfcc1V is difficult to interpret, we selected H1-A3 and conducted the analysis on it. H1-A3 represents the harmonic difference which is defined as the ratio of energy of the first harmonic (H1) to the energy of the highest harmonic in the third formant range (A3). FIG.\ref{fig:H1A3} and TABLE \ref{tab:stats} demonstrate the H1-A3 distributions and statistical summary across the age range for both sexes. It is noteworthy that the emergence of a substantial sex difference happens at the age of 10, with girls showing decreasing values in H1-A3 with advancing age while there is no clear age trend in boys. It is known that H1-A3 is correlated with the source spectral tilt \citep{hanson1999glottal}, where lower spectral tilt values indicate more abrupt glottal closures and higher spectral tilt values reflect less non-simultaneous closure. The results of our study indicate that the source spectral tilt is sex dependent in children older than 10 years. 

\subsubsection{\label{sec:4:4:4} VTL}
The factor of VTL is composed of 5 features across the age range, including avgF, $\Delta$f, pF, mff and fitch\_vtl. As these features were all highly correlated, we selected the Fitch formant estimate (fitch\_vtl) as the representative feature of this factor. FIG.\ref{fig:fitchvtl} and TABLE \ref{tab:stats} show the fitch\_vtl distribution and effect size across the age range for both sexes. As expected, fitch\_vtl, as an estimator of VTL, shows a substantial sex difference, but restricted to older children (aged 13-15), likely a consequence of puberty. Specifically, boys show a significant increase in fitch\_vtl after 13 years, while fitch\_vtl does not vary significantly with age in girls. This detailed pattern of results supports adding the VTL estimators to the eGeMAPS feature set to improve sex classification for children, as we demonstrated in Section \ref{sec:4:1}.

\subsubsection{\label{sec:4:4:5} Spectral flux}
Spectral flux (SF) represents a quadratic, normalised version of the simple spectral difference between the spectra of two consecutive speech frames. This factor is composed of three features, including the SF of all regions (SF mean), voiced regions (SF V), and unvoiced regions (SF UV). SF mean was selected as representative of these three features, as they are highly correlated. FIG.\ref{fig:sfmean} and TABLE \ref{tab:stats} show the distribution and the statistical summary of SF mean values for both sexes across the age range. 

We found that boys and girls showed different age trends for SF mean, with SF mean being consistently low for boys up until age 9, then increasing from age 10 to a consistently higher level, whereas for girls this transition occurred late, with SF mean levels remaining relatively low across ages 5-13 before increasing substantially for ages 14 and 15. These trends contributed to substantial sex differences favouring boys in the age range 9-13 followed by a reversal of the sex difference at age 15. SF mean is one indicator of the timbre of an audio signal, which means the timbre of children’s voices experience a significant change for boys at 9 years and for girls at 14 years. The sex difference favouring girls at age 15 is consistent with outcomes of a study investigating SF in discriminating adult male and female voices \citep{ghosal2014automatic}. 

\subsubsection{\label{sec:4:4:6} Unvoiced features}
The unvoiced features (UV) include the features of aR, hI and spectral slopes 0-500Hz of the unvoiced segments (aR UV, hI UV and slpUV0-500). Unvoiced speech is a type of speech without any periodic nature, but it is different to silence. During the production of unvoiced speech, the air exhaling out of the lungs through the trachea is not interrupted by the vibrating vocal folds. The air flow could be completely or narrowly obstructed by total or partial closure of somewhere along the length of the vocal tract, which results in stop or frication excitation and excites the vocal tract system to produce unvoiced speech. The typical examples of voiced signals are /a/, /e/, /i/, /u/, /o/; while the typical examples of unvoiced signals include stop consonants /p/, /t/, /k/ and some unvoiced fricative consonants /f/, /$\phi$/, /\textesh/. 

As the four inclusive features are highly correlated, we selected aR\_UV as a representative of this cluster. FIG.\ref{fig:aruv} and TABLE \ref{tab:stats} present the distributions and the statistical summary of aR\_UV.  In the present work, we found that the sex difference appears at ages 14 and 15, with the aR\_UV decreasing in girls while increasing in boys. 

\subsubsection{\label{sec:4:4:7} Loudness}
The factor of loudness (LD) includes LD mean, LD p20, LD p50, LD p80 and SL. Loudness is an estimate of perceived signal intensity from an auditory spectrum. Due to the high correlations among the inclusive features, we analysed the LD mean as a representative of the factor. FIG.\ref{fig:ldmean} and TABLE \ref{tab:stats} shows the loudness distribution and effect size across the age range for both sexes.  It provides evidence of girls sounding softer than boys at age 10 but louder than boys at age 15. Most previous studies used the F0 to estimate the highness or lowness of a tone perceived by the ear or used decibel (dB) to measure vocal intensity, and assess sex differences. The loudness related features used in the current study, different from F0 and vocal intensity, are a set of features that approximate humans’ non-linear perception of sound by applying an auditory spectrum in the Perceptual Linear Prediction\citep{hermansky1990perceptual}. Limited studies were found to compare with our findings regarding the sex differences in loudness varying with age, given that the loudness features are relatively new and not well understood.

\section{\label{sec:5} Conclusions}
This study investigated a broad range of acoustic features and proposed a hierarchical clustering-based machine learning model for the sex classification of children and characterisation of the salient acoustic factors. The set of models provided promising results in predicting sexes in children aged from 5 to 15 years, by using the combination feature set of eGeMAPS and VTL estimators. We reported greater success in predicting sexes by using spontaneous speech rather than scripted speech, especially for younger children. Smaller sets of independent acoustic factors were then obtained after the application of hierarchical clustering for children of each year, which showed equivalent performance in sex classification  to performance when using the full set of features. The independent acoustic factors were used for salient acoustic factors characterisation. 

We then analysed and discussed the important acoustic factors regarding the sex differences and age variation. We found that, for young children, all acoustic factors contributed to the sex classification equivalently. In other words, there are no noticeable acoustic factors that contribute substantially to distinguishing the voices of young girls and boys. For older children, sex differences become more prominent. Among the important factors, we found F0, HNR, Fn and H1-A3 showed significant sex differences for children aged above 10 years; VTL estimators are critical contributors in sex classification in children aged above 13 years;  SF has medium to large effect size in differentiating sexes for children at 10 and 13 years old; unvoiced features differentiate the two sexes for children aged above 14 years old; and girls and boys have different loudness in their speech when they are at 10 and 15 years old.

The limitations of this study are that: (1) the data were not strictly balanced between sexes for each year. For instance, in terms of the spontaneous samples, the number of boys was almost twice the number of girls at the age of 6 (1190:668) and 14 (1171:596), and the ratio was 2.86:1 between boys and girls at age 15 (1388:486). This imbalance could cause bias in sex classification, where the samples of the minority class could be misclassified to be the majority class; and (2) though we found the classification models were more accurate for spontaneous speech than for scripted speech, the sample size of spontaneous speech is considerably smaller than the sample size of scripted speech, especially for younger children (see TABLE \ref{tab:dataset}).  Future studies could consider deploying a more sex-balanced dataset and a larger dataset with more spontaneous speech.

\begin{acknowledgments}
The authors would like to thank Khaldoun Shobaki, John-Paul Hosom, and Ronald A. Cole for making the OGI Kids’ Speech Corpus available.
\end{acknowledgments}

\appendix*
\section{Procedures of Clustering}

\begin{figure}[hb]
\centering\vspace{8pt}
\includegraphics[width=0.4\linewidth]{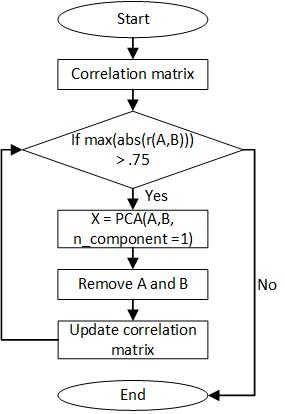}
\caption{Scheme of correlation-matrix-based hierarchical clustering}
\label{fig:cluster}
\end{figure}
Step 1. Each feature of the original feature set was assigned to be the simplest object of size 1. Assuming the original feature set was composed of m features, a correlation matrix of size m x m was first obtained. During the following steps, m would be reduced by 1 per iteration.

Step 2.	If the maximum absolute value of the correlation coefficients between two objects A and B, max(abs(r(A,B)), was larger than the cut-off value (0.75), the decision was made that the objects A and B were highly correlated and should be merged in the subsequent steps. Otherwise, all the objects in the current pool were regarded to be independent and there was no severe multicollinearity. The objects A or B could be clusters of any size larger than 1. There is no universal agreement on the cut-off value of the correlation coefficient in multicollinearity diagnostic. Investigators have used a cut-off value of 0.5 or greater \citep{donath2012predictors}. In the present study, we set the cut-off value to be 0.75.

Step 3.	Once the objects A and B were determined to be highly correlated, a one-component principal component analysis (PCA) was applied on all features wrapped in A and B to generate a new object X to represent A and B. For instance, if A was composed of 2 features, and B was composed of 3 features, a PCA would be applied on these 5 inclusive features. We regarded the newly generated object X as representative of the 5 features.

Step 4.	The two objects A and B were then removed from the pool that generated the correlation matrix in step 1 and replaced with X. The current pool would then be composed of m-1 objects, including the one newly generated object. The correlation matrix was then updated to be of size (m-1)$\times$(m-1). 

Step 5.	Steps 2 to 4 were repeated until all the absolute values of the pairwise correlation coefficients were lower than 0.75.
\bibliography{Manuscript}
\end{document}